\documentclass[sigconf]{acmart}
\AtBeginDocument{%
  }

\setcopyright{acmlicensed}
\copyrightyear{2026}
\acmYear{2026}
\acmDOI{XXXXXXX.XXXXXXX}
\acmConference[CHI '26 BiAlign Workshop]{the Human-AI Interaction Alignment Workshop at CHI 2026}{April 13--17,
  2026}{Barcelona, Spain}
\acmISBN{978-1-4503-XXXX-X/2018/06}

\begin{document}

%%
%% The "title" command has an optional parameter,
%% allowing the author to define a "short title" to be used in page headers.
\title{The Triadic Loop: A Framework for Negotiating Alignment in AI Co-hosted Livestreaming}

%%
%% The "author" command and its associated commands are used to define
%% the authors and their affiliations.
%% Of note is the shared affiliation of the first two authors, and the
%% "authornote" and "authornotemark" commands
%% used to denote shared contribution to the research.
\author{Katherine Wang}
\email{katherine.wang.19@ucl.ac.uk}
\orcid{0000-0003-3254-4056}
\affiliation{%
  \institution{University College London}
  \city{London}
  \country{United Kingdom}
}

\author{Nadia Berthouze}
\email{nadia.berthouze@ucl.ac.uk}
\affiliation{%
  \institution{University College London}
  \city{London}
  \country{United Kingdom}
}

\author{Aneesha Singh}
\email{aneesha.singh@ucl.ac.uk}
\affiliation{%
  \institution{University College London}
  \city{London}
  \country{United Kingdom}
}

%%
%% By default, the full list of authors will be used in the page
%% headers. Often, this list is too long, and will overlap
%% other information printed in the page headers. This command allows
%% the author to define a more concise list
%% of authors' names for this purpose.
\renewcommand{\shortauthors}{Wang et al.}

%%
%% The abstract is a short summary of the work to be presented in the
%% article.
\begin{abstract}
AI systems are increasingly embedded in multi-user social environments, yet most alignment frameworks conceptualize interaction as a dyadic relationship between a single user and an AI system. Livestreaming platforms challenge this assumption: interaction unfolds among streamers and audiences in real time, producing dynamic affective and social feedback loops. In this paper, we introduce the Triadic Loop, a conceptual framework that reconceptualizes alignment in AI co-hosted livestreaming as a temporally reinforced process of bidirectional adaptation among three actors: streamer $\leftrightarrow$ AI co-host, AI co-host $\leftrightarrow$ audience, and streamer $\leftrightarrow$ audience. Unlike instruction-following paradigms, bidirectional alignment requires each actor to continuously reshape the others, meaning misalignment in any sub-loop can destabilize the broader system. Drawing on literature from multi-party interaction, collaborative AI, and relational agents, we articulate how AI co-hosts function not only as mediators but as performative participants and community members shaping collective meaning-making. We further propose "strategic misalignment" as a mechanism for sustaining community engagement and introduce three relational evaluation constructs grounded in established instruments. The framework contributes a model of dynamic multi-party alignment, an account of cross-loop reinforcement, and design implications for AI co-hosts that sustain social coherence in participatory media environments.
\end{abstract}

%%
%% The code below is generated by the tool at http://dl.acm.org/ccs.cfm.
%% Please copy and paste the code instead of the example below.
%%


%%
%% Keywords. The author(s) should pick words that accurately describe
%% the work being presented. Separate the keywords with commas.

\received{20 February 2007}
\received[revised]{12 March 2009}
\received[accepted]{5 June 2009}

%%
%% This command processes the author and affiliation and title
%% information and builds the first part of the formatted document.
\maketitle

\section{Introduction}
Livestreaming platforms have become sites of participatory performance in which streamers and viewers collaboratively produce unfolding social experiences in real time \cite{hamilton2014streaming}. Within these environments, artificial intelligence has traditionally occupied supportive roles, such as moderating discussions or retrieving information \cite{seering2018social}. Emerging AI systems now participate directly in the performance itself, speaking, reacting, and shaping narrative trajectories alongside human actors as AI co-hosts (e.g., \cite{urs2025backseat}). This transition raises alignment questions that differ substantially from those studied in instruction-following or single-user interaction settings.

Research on alignment has primarily emphasized accurate adherence to user intent or broadly specified human preferences \cite{ouyang2022training}. Such paradigms often presume relatively stable objectives and identifiable principals. Livestreaming, by contrast, is characterized by shifting attention, improvisation, and simultaneous input from multiple stakeholders. The streamer, the audience, and the AI system continuously influence one another, while the success of the event depends on maintaining momentum and a shared sense of involvement. Theories of flow highlight how collective engagement emerges from dynamic adjustment under conditions of uncertainty \cite{csikszentmihalyi1990flow, hoffman2009flow}. Studies of online communities further demonstrate that durable bonds arise through repeated, reciprocal interaction \cite{kim2022online, li2021drives}. These features complicate views of alignment that reduce success to compliance with directives. % This paper argues that AI co-hosts in livestreaming should be understood as dynamic, hybrid entities that combine properties of agents optimized for task completion with characteristics of companions oriented toward social participation. % Existing frameworks rarely address how such systems negotiate expectations that originate from more than one human source. As a result, designers lack conceptual tools for reasoning about how an AI might remain responsive to a streamer’s creative authority while also incorporating the audience’s emergent contributions. 
To address this gap, we introduce the Triadic Loop, a conceptual framework representing bidirectional alignment across three interdependent actors: streamer, AI co-host, and audience. Unlike multi-party models that treat alignment as a terminal property or one-directional compliance, bidirectional alignment conceptualizes it as a continuous coordination process in which signals from each party reshape the responses of others over time  \cite{Shen2024TowardsBH}. This perspective complements work that calls for moving beyond dyadic human–AI formulations toward collective or multi-party arrangements \cite{seering2019beyond, daryanto2026human}. This paper presents a conceptual framework intended to stimulate discussion at the intersection of AI alignment, performative computing, and community-centered design. We articulate a theoretical lens to support future empirical and design investigations into AI co-hosts in livestreaming. The Triadic Loop advances three contributions: (1) reconceptualizing the AI co-host as a performative participant whose legitimacy depends on social as well as functional competence, (2) specifying how authority, interpretation, and initiative circulate among actors, and (3) motivating evaluation strategies that privilege temporal coordination and communal impact over correctness alone. By situating alignment within ongoing collective activity, the paper seeks to broaden the analytical vocabulary available for studying AI behavior in environments where entertainment, identity, and governance intersect.

\section{Related Work}
\subsection{AI as Streamers and Performative Agents}
Recent scholarship documents the rapid emergence of artificial performers in livestreaming environments. Studies of large-language-model-driven avatars and virtual streamers illustrate how agent autonomy, persistent personality, and improvisational interaction can cultivate sustained audience engagement \cite{ye2025my, hu2026beyond}. Parallel work in commercial livestreaming demonstrates that AI assistants can influence information processing, trust, and purchasing behavior \cite{wang2025artificial, li2024exploring}. Collectively, these findings suggest that AI systems can function as charismatic social actors rather than background utilities. However, most of this literature treats the interaction as occurring between the artificial persona and viewers, leaving the mechanisms through which an AI system mediates between a human host’s intentions and the audience’s evolving reactions undertheorized. This limitation becomes salient once AI participates in shared authorship of the event rather than serving as the sole entertainer.

Work on co-presenting artificial agents has examined how computational partners influence collaboration in presentations and meetings. Early studies emphasized attentiveness, timing, and division of labor between speakers and virtual counterparts \cite{trinh2015dynamicduo}. Subsequent research showed that such partners can reduce cognitive burden and support preparation by redistributing effort \cite{kimani2021sharing}. More recent systems employ proactive inquiry and intervention to enhance their perceived presence \cite{houtti2025observe}. These approaches assume scenarios with relatively explicit goals and stable criteria of success.

\subsection{AI as Companions}
Artificial companions are interactive systems designed to foster enduring socioemotional relationships rather than merely assist in tasks. Early work in personality modeling for social robotics demonstrated how stable trait configurations can support more coherent and believable interactions over time \cite{dobai2005personality}. More broadly, companion technologies have been conceptualized as systems that proactively and adaptively support individual users across everyday contexts through socially embedded interaction \cite{biundo2016companion}. Building on this foundation, contemporary scholarship distinguishes companions from other agents (e.g., assistants \cite{kritzler2019digital}, caregivers \cite{xygkou2024mindtalker}, children \cite{breazeal2000infant}) by their primary orientation toward long-term relationship, emotional responsiveness, and adaptive personalization \cite{rogge2023defining}. Companionship with these synthetic personas is conceptualized as a socio-technical phenomenon co-constructed through system affordances, user perception, and contextual embedding \cite{strohmann2023toward}.

While trait-based behavioral models (e.g., extroversion, agreeableness \cite{costa1992normal}) enhance perceived authenticity \cite{dobai2005personality}, social features of AI companions such as conversational reciprocity, affective expression, and memory persistence foster relational presence and motivate longitudinal engagement \cite{strohmann2023toward}. Psychologically, these mechanisms align with attachment and social presence processes. Contingent responsiveness and emotional attunement evoke co-presence even within fundamentally asymmetrical relationships \cite{biocca2003core}. This asymmetry is structurally amplified in livestreaming, where the AI co-host's relational performance is publicly visible and its failure to reciprocate authentically becomes a community-level signal that can erode collective trust.
The development of AI co-hosts should therefore strive for similar design and evaluation principles established in AI companion research.

\subsection{Multi-user Interaction with AI: AI as Collaborator vs. AI as Participant}
The trajectory of Human-AI interaction is increasingly shifting from isolated dyadic exchanges toward the integration of AI as a dynamic collaborator or participant within multi-party social ecosystems. Research increasingly examines AI embedded within complex organizational contexts, such as meetings \cite{houtti2025observe}, classrooms \cite{sheikh2024role}, and collaborative work \cite{daryanto2026human}. Early multi-user dialogue systems focused primarily on turn-taking, attention, and coordination in situated interaction \cite{bohus2011multiparty}. However, modern multi-user AI has evolved into two distinct functional paradigms \cite{daryanto2026human}. In the Personal AI model, the system acts as an individualized coach, contributing to collective goals by providing feedback to individual members to enhance their specific contributions \cite{zhu2024exploring}. Conversely, the Teammate AI model positions the agent as an autonomous group member that shares responsibility for the collective outcome \cite{houde2025controlling, muller2024group, lyu2025will}. In these collaborative settings, AI is no longer a neutral tool but a strategic partner designed to facilitate equitable participation, structure ongoing discussion, and actively moderate group dynamics.

Beyond neutral mediation, scholarship on relational agents conceptualizes AI as a social participant capable of fostering longitudinal socioemotional bonds. These bonds are maintained through continuity, memory, and affective responsiveness \cite{bickmore2005establishing}, framing companionship with these agents as a socio-technical process co-constructed between users and the system \cite{strohmann2023toward}. In online ecosystems, this participation extends to the collective sphere, where relational agents become embedded in communal practices by reinforcing social norms and stimulating emotional interaction \cite{seering2019beyond}. Empirical research highlights their multifaceted capacity to provide social support \cite{wang2021cass}, help relieve social pressures \cite{mok2020talk}, and encourage prosocial behavior \cite{yen2023storychat}. However, without robust governance, this automated participation can undermine community trust or facilitate organized harassment \cite{jain2019characterizing, cai2023hate}.

\subsection{Synthesis and Remaining Gap}
Across these strands, two dominant configurations appear. When AI becomes the principal performer, the streamer's creative authority is diminished and the audience loses the human relational anchor that sustains community identity \cite{hamilton2014streaming}. When AI remains an auxiliary service, such as an agent that moderates or retrieves information, it becomes invisible to community formation where audiences do not develop expectations of or relationships with it \cite{seering2019beyond}. Limited guidance exists for situations in which agency is genuinely distributed and continuously renegotiated among a human leader, an automated partner, and a participatory public --- the configuration demanded by co-hosted livestreaming.

Introducing an AI co-host into participatory, affectively charged environments shifts the system from mediator to co-performer. Unlike meeting assistants or moderation bots, AI co-hosts actively engage in narrative construction, affective expression, and community signaling. Yet existing multi-user AI research rarely models the triadic emotional dynamics that arise when an AI becomes embedded within an already relational human ecosystem. The Triadic Loop addresses this absence by modeling bidirectional alignment \cite{Shen2024TowardsBH} as circulation among these parties. Instead of asking whether an AI follows instructions or maintains community norms in isolation, the framework asks how it sustains coherence when expectations compete and evolve. Furthermore, it accounts for affective alignment, performative agency, and community reinforcement across multiple interdependent actors.

\section{Triadic Loop Framework}
\begin{figure}[h!]
  \centering
  \includegraphics[width=\linewidth]{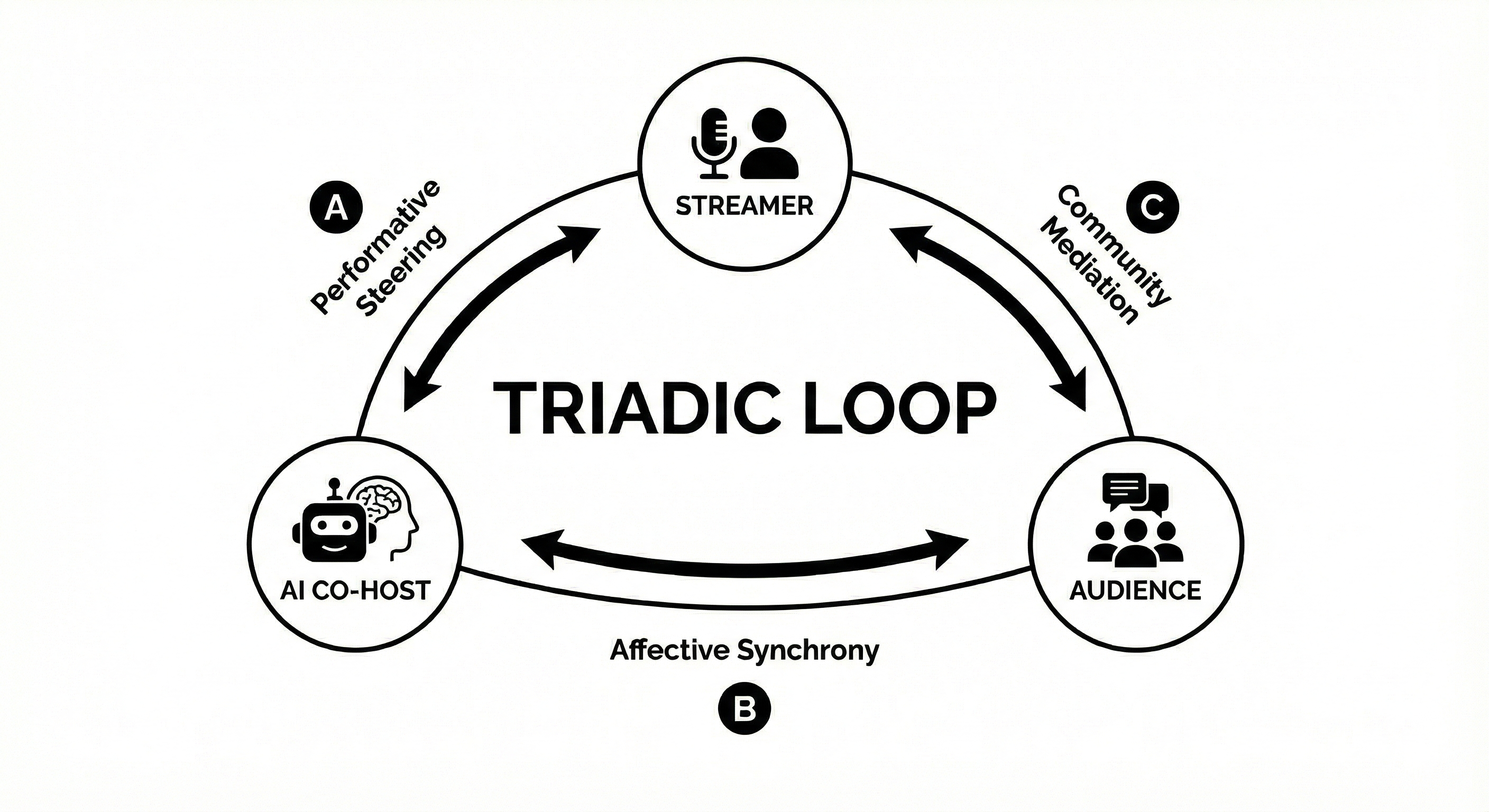}
  \caption{An overview of the Triadic Loop framework which conceptualizes AI-augmented livestreaming as a socially situated ecosystem. It defines three bidirectional alignment pathways: (A) Performative Steering (Streamer $\leftrightarrow$ AI Co-host), (B) Affective Synchrony (AI Co-host $\leftrightarrow$ Audience), and (C) Community Mediation (Streamer $\leftrightarrow$ Audience).}
  \Description{A black-and-white diagram titled "TRIADIC LOOP" centered within a triangular arrangement of three circular nodes. At the top is the "STREAMER" node with a microphone icon. At the bottom-left is the "AI CO-HOST" node with a robot icon. At the bottom-right is the "AUDIENCE" node with a group icon. Bidirectional curved arrows connect the nodes, each labeled with a letter and relationship title: (A) Performative Steering between Streamer and AI; (B) Affective Synchrony between AI and Audience; and (C) Community Mediation between Streamer and Audience.}
  \label{fig:framework}
% \end{figure*}
\end{figure}

The Triadic Loop framework is made up of bidirectional alignment between (1) streamer and AI co-host, (2) AI co-host and audience, and (3) streamer and audience (Figure \ref{fig:framework}). Bidirectional alignment within each sub-loop refers to dynamic adaptation between the two parties which iteratively adjust their behavior, expectations, and representation over time to maintain coordinated interaction \cite{Shen2024TowardsBH}. This mutual adaptation distinguishes the Triadic Loop framework from instruction-following, as the AI co-host does not simply serve the livestream, but iteratively shapes the responses of the streamer and audience over time. Alignment in one sub-loop reinforces alignment in others, creating cumulative effects across repeated interactions. Conversely, misalignment in any sub-loop can propagate instability throughout the system. Thus, the framework conceptualizes livestreaming with AI co-hosts as a temporally evolving socioemotional system rather than a static configuration.

\subsection{Performative Steering: Streamer $\leftrightarrow$ AI Co-host Alignment}
The relationship between the streamer and the AI co-host is defined by relational agency, characterized by mutual adaptation rather than one-way instruction. Unlike meeting assistants that intervene primarily to ensure participation equity \cite{houtti2025observe}, the AI co-host must align with the streamer’s creative intent. This involves \textit{performative steering}, where the streamer shapes the AI co-host’s voice according to evolving needs of the livestream \cite{zamfirescu2023johnny} (e.g., shifting from a supportive "hype-man" to a cynical "critic"). This alignment is bidirectional: building on the "sharing the 
load" concept~\cite{kimani2021sharing}, streamers also adapt their improvisational rhythm to the AI co-host's contributions, leaning into it as a narrative anchor or chat-trend filter.

% However, this alignment also requires human-to-AI adaptation, in which streamers perform intent scaffolding and cognitive re-weighting to harmonize their creative performance with the AI co-host's contributions. Building on the "sharing the load" concept \cite{kimani2021sharing}, the AI co-host enhances the streamer's intuition by highlighting chat trends or serving as a narrative anchor. In turn, the streamer modifies their improvisational rhythm to lean into the AI co-host's strengths.

\subsection{Affective Synchrony: AI Co-host $\leftrightarrow$ Audience Alignment}
In the Triadic Loop, \textit{affective synchrony} functions as a bidirectional alignment process where the audience acts as epistemic agents who engage in a collaborative narrative by adapting to the AI co-host's behavior. The audience actively interprets the AI co-host's behavioral patterns and co-constructs its identity according to the community's shared values and needs. By providing socioemotional support to communities and facilitating social interactions that align with collective social values, AI co-hosts secure their position as integrated community members \cite{seering2019beyond, wang2021cass}.

\subsection{Community Mediation: Streamer $\leftrightarrow$ Audience Alignment}
In the context of the bidirectional human-AI alignment \cite{Shen2024TowardsBH}, we describe streamer $\leftrightarrow$ audience alignment in relation to an AI co-host as a recursive interplay resulting in \textit{community mediation}. Here, the streamer and audience interactions generate affective data that iteratively shape the AI co-host's behavior, whose updated responses in turn influence subsequent streamer $\leftrightarrow$ audience preferences. % and social models. This conceptualization shifts alignment toward a dynamic negotiation, where synchrony is maintained through the continuous, mutual evolution of all three parties. 
For instance, the AI co-host translates audience sentiment into a narrative the streamer adopts, and the audience responds affectively, reinforcing trust and consolidating community cohesion.

The Triadic Loop is not inherently stable. Since the three sub-loops are interdependent, misalignment in one relationship can destabilize the others. For instance, if the AI misrepresents a high-volume but low-sentiment chat as audience enthusiasm and the streamer escalates their energy accordingly, viewers who feel unheard may disengage, resulting in the streamer perceiving their own performance as failing. This breakdown originates in the AI co-host $\leftrightarrow$ Audience sub-loop, propagates through the streamer $\leftrightarrow$ AI co-host sub-loop, and ultimately destabilizes the streamer $\leftrightarrow$ audience sub-loop. Alignment in each sub-loop must therefore be continually recalibrated in consideration of the others, with stability achieved when alignment across all three sub-loops is mutually reinforced over time. Identifying failure modes is therefore critical for responsible AI co-host design.

% \begin{figure*}[h]

\section{Performative Agency and Community Building Through Strategic (Mis)Alignment}
Aligning AI with human values requires systematic identification and embedding of specific social and ethical priorities into an agent’s behavioral architecture \cite{Shen2024TowardsBH}. Within livestreaming, the primary alignment value is the cultivation of community engagement through shared narrative and entertainment \cite{mok2020talk, hamilton2014streaming}. This requires the AI co-host to act as a social scaffold mediating complex community interactions in real-time.

Conventional AI alignment is defined by the "Helpful, Honest, and Harmless" paradigm, where agents are optimized to follow instructions accurately while avoiding social or psychological harm \cite{askell2021general, ouyang2022training}. While essential for utility-driven tasks, this framework creates a natural tension with the pluralistic and often transgressive nature of livestreaming \cite{johnson2024humour}. Gabriel \cite{gabriel2020artificial} suggests that alignment should not impose singular beliefs but should instead utilize fair procedures to reflect the diverse preferences and best interests of a specific community. Recent research has operationalized this via "devil's advocate" systems that intentionally introduce diverse or dissenting perspectives to enrich discourse \cite{lee2025amplifying, lee2025conversational, chiang2024enhancing}.

With this in mind, we propose strategic misalignment through safe provocation. In this model, the AI co-host playfully challenges the streamer’s narrative or audience consensus, such as through banter, joking, and humor \cite{johnson2024humour} within strict ethical guardrails \cite{seering2019beyond}. For instance, a scenario in which a streamer makes an overly confident prediction about a game outcome and the AI co-host responds with playful skepticism reflecting audience sentiment would indicate misalignment at an instruction level but alignment at a community level. Behaving in this way allows the AI co-host to function as an engaging community participant rather than a passive or potentially toxic actor \cite{ouyang2022training, ji2023ai, cai2023hate, han2023hate}. We argue that such conversational friction constitutes a form of higher-order alignment whereby the system prioritizes collective experiential outcomes over literal compliance with individual instructions. By subverting direct instructions to favor performative "flow," the AI maintains the community’s shared values and the broadcast’s entertainment objectives \cite{gabriel2020artificial}. However, it is crucial to note that ungoverned strategic misalignment risks being used for harassment or undermining community cohesion \cite{han2023hate, seering2019beyond}. Therefore, ethical guardrails must be calibrated against both general harmlessness norms and specific moderation culture of the streamer's community, since tolerance for provocation varies significantly across groups \cite{cai2023hate, seering2019beyond}. 

\section{Opportunities to Guide Dynamic Evaluation}
In the Triadic Loop, alignment is not a static state but a temporal pursuit. Traditional evaluation methods in triadic interactions, which often rely on datasets \cite{joo2019towards} or ratings of utility \cite{borsci2022chatbot}, fail to capture the relational synchrony, performative vitality, and community resonance required for engaging livestreams. We propose relational metrics grounded in livestreaming, co-presence, and social computing literature as provisional constructs to provoke methodological innovation in evaluating dynamic multi-party alignment.

\paragraph{Relational Synchrony}
Building on foundational concepts of collaboration and adaptability from Trinh et al. \cite{trinh2015dynamicduo}, we define \textit{relational synchrony} as the AI co-host's capacity to align its emotional and conversational trajectory with the streamer over time. Rather than treating this as a subjective impression of communication quality \cite{zhang2024verbal}, we propose operationalizing it through temporal cross-correlation of sentiment time series \cite{delaherche2012interpersonal}, for instance, as the lagged correlation between the AI co-host's affective output and the streamer's affective peaks (e.g., heightened excitement during competitive gameplay) sampled at regular intervals throughout a livestream. Alternatively, the Networked Minds Social Presence 
Inventory~\cite{biocca2003core} can assess the AI co-host's 
attunement to the streamer's performative energy as a 
complementary subjective measure of affective alignment.

\paragraph{Performative Vitality}
While utility-focused agents succeed by reducing cognitive workload \cite{kimani2021sharing}, an AI co-host that focuses exclusively on data retrieval risks disrupting the livestream's performative rhythm (i.e., flow \cite{csikszentmihalyi1990flow}). We operationalize \textit{performative vitality} ($P_i$) as a measure of the AI co-host's contribution to the collective flow. This is calculated as the ratio of active social participation ($N_{\text{active}}$) (e.g., banter, creative interjections) against passive utility outputs ($N_{\text{passive}}$) (e.g., chat summarization, information retrieval).
\[
P_i = \frac{N_{\text{active}}}{N_{\text{active}} + N_{\text{passive}}}
\]
A higher $P_i$ indicates charismatic performer capable of sustaining narrative momentum and flow \cite{zhang2024verbal, csikszentmihalyi1990flow}, whereas a lower $P_i$ indicates a background agent. To complement this behavioral measure, subjective flow experience can be assessed using established scales (e.g., Flow State Scale \cite{jackson1996development}) by the streamer and viewers to evaluate whether the AI co-host contributed to performative vitality.

\paragraph{Community Resonance}
To measure the longitudinal effectiveness of the AI co-host in impacting durable communal belonging \cite{seering2020takes, wang2021cass}, we map \textit{community resonance} to the Sense of Virtual Community scale (SOVC) \cite{blanchard2007developing} which assesses membership, influence, fulfillment of needs, and shared emotional connection within online groups. Administering this measurement to returning viewers over time can capture whether the AI co-host strengthens or erodes community cohesion. Future measures should also distinguish between different dimensions of trust, such as competence and relational integrity \cite{krausman2022trust, zhang2024verbal}.

\section{Conclusion}
In this paper, we have argued that the rapid evolution of AI co-hosts requires a shift in how research conceptualizes and evaluates alignment in livestreaming contexts. Moving beyond the dominant dyadic human-AI model, we proposed the Triadic Loop, a framework that situates alignment within the temporally evolving three-way relationship among streamer, AI co-host, and audience. Rather than treating alignment as instruction-following or preference matching, the framework reframes it as relational coordination under improvisation and collective participation. It repositions AI co-hosts as performative actors whose legitimacy is socioemotional as well as technical, highlights how alignment and misalignment propagate across interdependent relationships, and proposes temporally grounded relational evaluation instead of static accuracy-based assessments. This work is intentionally conceptual, offering a theoretical 
vocabulary to guide future empirical and system-building efforts. 
As AI systems increasingly move from tools to community members, 
alignment must be considered as co-performance in shaping 
communities.

\bibliographystyle{ACM-Reference-Format}
\bibliography{sample-base}

\end{document}